\documentclass{article}
\usepackage{spconf,amsmath,amssymb,graphicx,hyperref,bm,standalone,tikz,subcaption,pgfplots}
\usetikzlibrary{calc, trees, positioning, arrows, shapes, shapes.geometric, fit, backgrounds, patterns}
\pgfplotsset{compat=1.17}
\def\L{{\cal L}}

\title{Towards Local Rate Control in Neural Video Compression}
\name{Marc Windsheimer, Simon Deniffel, and Andr\'e Kaup\thanks{The authors gratefully acknowledge that this work has been funded by the Deutsche Forschungsgemeinschaft (DFG, German Research Foundation) under project number 426084215.}\thanks{The authors gratefully acknowledge the scientific support and HPC resources provided by the Erlangen National High Performance Computing Center (NHR@FAU) of the Friedrich-Alexander-Universität Erlangen-Nürnberg (FAU). The hardware is funded by the German Research Foundation (DFG).}}
\address{\textit{Chair of Multimedia Communications and Signal Processing}\\\textit{Friedrich-Alexander-Universität Erlangen-Nürnberg}\\Cauerstr. 7, 91058 Erlangen, Germany\\\{marc.windsheimer, simon.deniffel, andre.kaup\}@fau.de}
\begin{document}
%\ninept 
%
\maketitle
\begin{abstract}
Local rate control is a key enabler to generalize image and video compression for dedicated challenges, such as video coding for machines.
While traditional hybrid video coding can easily adapt the local rate-distortion trade-off by changing the local quantization parameter, no such approach is currently available for learning-based video compression.
In this paper, we propose a hierarchical video compression network, which allows continuous local rate control on a pixel level to vary the spatial quality distribution within individual video frames.
This is achieved by concatenating a quality map to the input frame and applying a weighted MSE loss which matches the pixel-wise trade-off factors in the quality map.
During training, the model sees a variety of quality maps due to a constrained-random generation.
Our model is the first neural video compression network which can continuously and spatially adapt to varying quality constraints.
Due to the wide quality and bit rate range, a single set of network parameters is sufficient.
Compared to single rate point networks, which scale linearly with the number of rate points, the memory requirements for our network parameters remain constant at nearly the level of one single rate model.
The code and model are available at \url{https://github.com/FAU-LMS/LRC-DHVC}.
\end{abstract}
\section{Introduction}
\label{sec:intro}
Visual data, in particular video content, accounts to the vast majority of all internet traffic~\cite{sandvine2023}, and as resolutions and frame rates continue to rise, efficient compression becomes ever more critical.
Traditional video codecs rely on block-based transforms and handcrafted rate-distortion optimization, but struggle to improve compression performance without adding a significant amount of complexity.
With~\cite{lu2019}, neural video compression (NVC) has emerged as alternative, by learning compact latent representations in a end-to-end manner and jointly optimizing transformation, temporal modeling, and entropy coding from training data.
However, most existing NVC schemes are either limited to single rate points or manage rate allocation solely using global Lagrangian multipliers across entire sequences or on a frame-wise level.

In contrast, local rate control aims to distribute the rate efficiently across both, space and time, tailoring the compression to satisfy scenario-specific quality and rate constraints.
In conventional codecs, local rate control is performed using spatial QP adaptation and block-level bit allocation.
Translating these ideas to NVC, however, poses unique challenges, since non-stationary latent statistics have to be covered by a single end-to-end compression framework.
Local rate control in neural video compression allows wide adoption for many use cases, e.g.  for privacy-preserving compression or video coding for machines.

\section{Related Work}
\begin{figure*}[t]
	\centering
		\begin{tikzpicture}[
		dataBlob/.style={rectangle, anchor = center},
		Layer/.style={rectangle, draw, anchor=center, minimum width=1cm, minimum height=0.75cm, rounded corners},
		Misc/.style={rectangle, draw, anchor = center, fill=black!20},
		Code/.style={, draw, anchor = center, fill=black!20},
		bg/.style={rectangle, anchor = center, fill=\colorBG, anchor = north west,},
		myArrow/.style={-stealth},
		label/.style={rectangle, anchor = center},
		operator/.style={circle ,draw, anchor = center, minimum size = 0.2cm, inner sep = 0pt},
		dot/.style={circle, fill, minimum size=3pt, inner sep=0pt}
		]
		
		\node (x) [dataBlob] at (0, 0) {\small$\bm{x}_t$};
		\node (m) [dataBlob, below of=x, node distance=0.66cm, color=red] {\small$\bm{m}_t$};
		\node (a1) [trapezium, draw, right of=x, node distance=1.25cm, yshift=-0.33cm, rotate=-90, minimum width=1.7cm, fill=gray!20] {};
		\node (dots1) [dataBlob, right of=a1, node distance=1.25cm] {...};
		\node (a2) [trapezium, draw, right of=dots1, node distance=1.25cm, rotate=-90, minimum width=1.2cm, fill=gray!20] {};
		\node (dummy1) [dataBlob, right of=a2, node distance=1.75cm] {};
		\node (ch1) [Layer, below of=dummy1, node distance=2cm, align=center, dashed, draw=red, fill=red!10, text=red] {\small Latent\\\small Block};
		\node (prior1) [dataBlob, below of=ch1, node distance=1.1cm] {\small$\bm{F}^L_{<t}$};
		\node (dots2) [dataBlob, right of=dummy1, node distance=1.75cm] {...};
		\node (a3) [trapezium, draw, right of=dots2, node distance=1.25cm, rotate=-90, minimum width=0.7cm, fill=gray!20] {};
		\node (dummy2) [dataBlob, right of=a3, node distance=1.5cm] {};
		\node (ch2) [Layer, below of=dummy2, node distance=2cm, align=center, dashed, draw=red, fill=red!10, text=red] {\small Latent\\\small Block};
		\node (bias) [dataBlob, right of=ch2, node distance=1.75cm] {};
		\node (bias_f) [dataBlob, above of=bias, node distance=0.35cm] {\small $\bm{f}^0_t$};
		\node (bias_d) [dataBlob, below of=bias, node distance=0.35cm] {\small $\bm{d}^0_t$};
		\node (prior2) [dataBlob, below of=ch2, node distance=1.1cm] {\small$\bm{F}^1_{<t}$};
		\node (s3) [trapezium, draw, below of=a3, node distance=2cm, rotate=-90, minimum width=0.7cm, fill=gray!20] {};
		\node (dots3) [dataBlob, below of=dots2, node distance=2cm] {...};
		\node (s2) [trapezium, draw, below of=a2, node distance=2cm, rotate=-90, minimum width=1.2cm, fill=gray!20] {};
		\node (dots4) [dataBlob, below of=dots1, node distance=2cm] {...};
		\node (s1) [trapezium, draw, below of=a1, node distance=2cm, rotate=-90, minimum width=1.7cm, fill=gray!20] {};
		\node (xhat) [dataBlob, left of=s1, node distance=1.25cm] {\small$\bm{\hat{x}}_t$};

		\draw[myArrow] (x) -- ([yshift=0.33cm] a1.south);	
		\draw[myArrow, red] (m) -- ([yshift=-0.33cm] a1.south);
		\draw[myArrow] (a1) -- (dots1);
		\draw[myArrow] (dots1) -- (a2);
		\draw[myArrow] (a2) -- (dots2);
		\draw[myArrow] (a2) -| (ch1) node[pos=0.5, dot] {} node[pos=0.5, above] {\small$\bm{r}^L_t$};
		\draw[myArrow] (dots2) -- (a3);
		\draw[myArrow] (a3) -| (ch2) node[pos=0.5, above] {\small$\bm{r}^1_t$};
		\draw[myArrow] ([xshift=-0.1cm,yshift=0.1cm] bias.west) -- ([yshift=0.1cm] ch2.east);
		\draw[myArrow] ([xshift=-0.1cm,yshift=-0.1cm] bias.west) -- ([yshift=-0.1cm] ch2.east);
		\draw[myArrow] ([yshift=0.1cm] ch2.west) -- ([yshift=0.1cm] s3.north) node[pos=0.5, above] {\small$\bm{f}^1_t$};
		\draw[myArrow] ([yshift=-0.1cm] ch2.west) -- ([yshift=-0.1cm] s3.north) node[pos=0.5, below] {\small$\bm{d}^1_t$};
		\draw[myArrow] ([yshift=0.1cm] s3.south) -- ([yshift=0.1cm] dots3.east);
		\draw[myArrow] ([yshift=-0.1cm] s3.south) -- ([yshift=-0.1cm] dots3.east);
		\draw[myArrow] ([yshift=0.1cm] dots3.west) -- ([yshift=0.1cm] ch1.east);
		\draw[myArrow] ([yshift=-0.1cm] dots3.west) -- ([yshift=-0.1cm] ch1.east);
		\draw[myArrow] (ch1) -- (s2) node[pos=0.5, below] {\small$\bm{d}^L_t$};
		\draw[myArrow] (s2) -- (dots4);
		\draw[myArrow] (dots4) -- (s1);
		\draw[myArrow] (s1) -- (xhat);
		
		\draw[myArrow] (prior1) -- (ch1);
		\draw[myArrow] (prior2) -- (ch2);
		
	\end{tikzpicture}
	\caption{Architectural overview of LRC-DHVC, our hierarchical video codec for local rate control using a pixel-granular quality map $\bm{m}_t$. The latent block incorporates temporal modeling via the temporal priors $\bm{F}^l_{<t}$, probability modeling, quantization and entropy coding. Initial contexts $\bm{f}^0_t$ and $\bm{d}^0_t$ are trainable parameters. Novel components are highlighted in red.}
	\label{fig:lrc_dhvc}
\end{figure*}
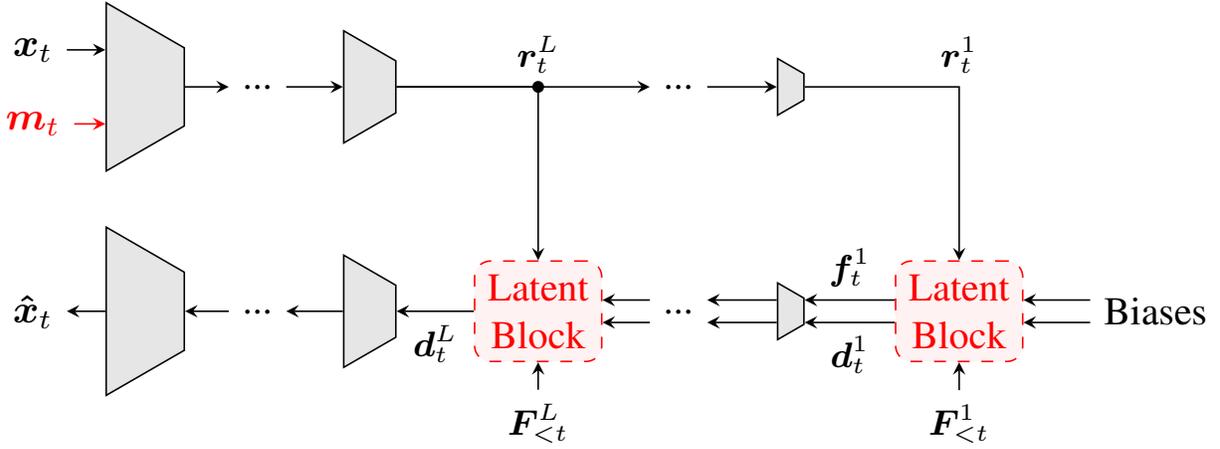
Rate control can be performed on multiple levels in the encoding process of video sequences.
Firstly, on the global level for the whole video sequence, secondly, on the frame level and finally, locally for the coding blocks within an individual frame.
Hybrid video codecs, such as HEVC~\cite{sullivan2012} and VVC~\cite{bross2021}, can perform rate control on all of these levels~\cite{wei2024}.

Due to the advances in neural image and video compression, rate control methods for neural compression networks are actively developed.
So far, rate control algorithms for neural compression mainly focus on global and frame level optimization.
Some variable rate methods for image compression train adaptive compression networks and are controlled by a global quality parameter during inference \cite{yin2022,iwai2024}.
Other approaches are controlled by specifying a target bit rate to control the rate allocation \cite{xue2024,zhang2025}.
In~\cite{li2022} the ideal Lagrangian $\lambda_i$ for the i-th frame is derived by modeling the interdependence of rate, distortion, and the Lagrangian.
DCVC-FM~\cite{li2024} applies learned scaling factors in the encoder and decoder transformation, where the scaling factors are indexable for each individual frame. 
In~\cite{gu2025}, per-frame Lagrangians are derived by matching predicted \mbox{$R$-$\lambda$} and \mbox{$R$-$D$} relations to a desired target bit rate.

Other methods, like code editing~\cite{gao2022} or semi-amortized variational inference~\cite{xu2023}, optimize bit allocation with respect to the traditional global rate-distortion trade-off during inference.

While the granularity of rate control in conventional transform coding is limited by the transform block size, neural compression enables rate control for individual pixels.
Prior work uses the Spatial Feature Transform (SFT)~\cite{wang2018}, for local rate control~\cite{song2021} in neural \textit{image} compression.
By training the compression network on a quality map-dependent loss, it can learn to distribute bit rate and quality accordingly.
Since the quality map is only an input to the encoder, there is no need to send additional side-information on the quality map.
Instead, the network inherently learns to include the required quality information as part of the latent $\bm{\hat{y}}$ and hyperlatent $\bm{\hat{z}}$.

Latent space masking~\cite{fischer2022} is used in image coding for machines to shift the latent values towards their predicted mean.
While this shift leads to higher distortions, the likelihood of the latent symbols is increased, resulting in a lower rate.
Similarly, saliency-dependent compression on varying hierarchical levels~\cite{fischer2023} allows a blockwise variation of the rate-distortion trade-off.
In \cite{shindo2025}, a Delta-Gaussian entropy model is used in conjunction with a binary region-learning-loss.

In neural \textit{video} compression, existing spatial rate control methods are limited with respect to the adjustment options.
ROI-DVC~\cite{wu2024} performs a Region-of-Interest (ROI)-based coding using a binary ROI map of the video frame.
The frame and the ROI map are split into multiple patches.
For each individual patch, the ratio of ROI area to total patch area is derived and depending on the ratio, one out of three models is selected for encoding the patch.
As the different models have been trained with varying rate-distortion trade-offs, patches with a high ROI area are encoded with a higher quality and vice versa.

\begin{figure}[t]
	\centering
		\begin{tikzpicture}[
		dataBlob/.style={rectangle, anchor = center},
		Layer/.style={rectangle, draw, anchor=center, minimum width=1.7cm, minimum height=0.75cm, rounded corners},
		Misc/.style={rectangle, draw, anchor = center, fill=black!20},
		Code/.style={, draw, anchor = center, fill=black!20},
		bg/.style={rectangle, anchor = center, fill=\colorBG, anchor = north west,},
		myArrow/.style={-stealth},
		label/.style={rectangle, anchor = center},
		operator/.style={circle ,draw, anchor = center, minimum size = 0.2cm, inner sep = 0pt},
		dot/.style={circle, fill, minimum size=3pt, inner sep=0pt}
		]
		
		\node (r) [dataBlob] at (0, 0) {$\bm{r}^l_t$};
		\node (res1) [Layer, below of=r, node distance=1.75cm, yshift=-0.75cm, fill=gray!20] {ResBlock};
		\node (res2) [Layer, right of=res1, node distance=1.75cm, fill=gray!20] {ResBlock};
		\node (pred) [Layer, right of=res2, node distance=2.35cm, fill=gray!20] {Prediction};
		\node (conv1) [Layer, below of=res1, node distance=1.6cm, xshift=0.875cm, fill=gray!20] {Conv};
		\node (res3) [Layer, below of=conv1, node distance=1.1cm, fill=gray!20] {ResBlock};
		\node (conv2) [Layer, below of=res3, node distance=1.1cm, fill=gray!20] {Conv};
		\node (div) [operator, below of=conv2, node distance=1.1cm, draw=red] {\color{red}$\div$};
		\node (q) [rectangle, draw, below of=div, node distance=1.1cm, minimum width=0.7cm, minimum height=0.7cm, fill=gray!20] {Q};
		\node (ae) [rectangle, draw, right of=q, yshift=-1.1cm, minimum width=0.7cm, minimum height=0.7cm, fill=gray!20] {AE};
		\node (mul) [operator, right of=ae, node distance=0.9cm, draw=red] {\color{red}$\times$};
		\node (conv3) [Layer, right of=mul, node distance=1.25cm, fill=gray!20] {Conv};
		\node (a) [operator, right of=conv3, node distance=1.25cm] {$+$};
		\node (res4) [Layer, above of=a, node distance=8.6cm, fill=gray!20] {ResBlock};
		\node (res5) [Layer, below of=a, fill=gray!20] {ResBlock};
		\node (res6) [Layer, right of=res5, node distance=2cm, fill=gray!20] {ResBlock};
		\node (conv4) [Layer, above of=res6, node distance=1.1cm, fill=gray!20] {Conv};
		\node (conv5) [Layer, below of=pred, node distance=1.5cm, fill=gray!20, draw=red, text=red] {Conv};
		\node (ftl0) [dataBlob, above of=res4, node distance=1cm] {$\bm{f}^{l-1}_t$};
		\node (dtl0) [dataBlob, right of=ftl0, node distance=2.33cm] {$\bm{d}^{l-1}_t$};
		\node (ftt) [dataBlob, left of=ftl0, node distance=2.5cm] {$\bm{F}^l_{<t}$};
		\node (ftl1) [dataBlob, below of=res5, node distance=1.5cm] {$\bm{f}^l_t$};
		\node (dtl1) [dataBlob, below of=res6, node distance=1.5cm] {$\bm{d}^l_t$};
		
		\draw[myArrow] (r) -- (res1);
		\draw[myArrow] (ftt) |- (pred);
		\draw[myArrow] (ftl0) -- (res4);
		\draw[myArrow] (dtl0) -- ([xshift=0.33cm] conv4.north);
		\draw[myArrow] (res4) -- ++(0, -0.75) node[pos=1, dot] {} -| (res2);
		\draw[myArrow] (res4) -- ++(0, -0.75) -| (pred) node[pos=0.5, dot] {};
		\draw[myArrow] (res4) -- (a);
		\draw[myArrow] (res1) -- ++(0, -0.75) -| ([xshift=-0.25cm] conv1.north);
		\draw[myArrow] (res2) -- ++(0, -0.75) -| ([xshift=0.25cm] conv1.north);
		\draw[myArrow] (conv1) -- (res3);
		\draw[myArrow] (res3) -- (conv2);
		\draw[myArrow] (conv2) -- (div);
		\draw[myArrow] (div) -- (q);
		\draw[myArrow] (q) |- (ae) node [pos=0.5, left] {$\hat{\bm{z}}_t^l$};
		\draw[myArrow] (ae) -- (mul);
		\draw[myArrow] (mul) -- (conv3);
		\draw[myArrow] (conv3) -- (a);
		\draw[myArrow] (pred) -- ++(0, -0.75) node[pos=1, dot] {} -| (conv4);
		\draw[myArrow] (pred) -- (conv5);
		\draw[myArrow] (a) -- (res5);
		\draw[myArrow] (res5) -- ++(0, -0.75) node (dot1) [pos=1, dot] {} -- ++(1, 0) -- ++(0, 2.5) -| ([xshift=-0.33cm] conv4.north);
		\draw[myArrow] (conv4) -- (res6);
		\draw[myArrow] (res5) -- (ftl1);
		\draw[myArrow] (res6) -- (dtl1);
		
		\draw[myArrow] (conv5.south) -- ++(0, -0.75) node[pos=1, below] {$\bm{\mu}_t^l$};
		\draw[myArrow] ([xshift=0.5cm] conv5.south) -- ++(0, -0.75) node[pos=1, below] {$\bm{\sigma}_t^l$};
		\draw[myArrow, red] ([xshift=-0.5cm] conv5.south) -- ++(0, -0.3) -| (mul);
		\draw[myArrow, red] (mul) |- (div) node [pos=0.5, dot] {} node [pos=0.5, right] {$\bm{\omega}_t^l$};
		\draw[myArrow] ([xshift=0.25cm, yshift=-1.5cm] conv5.south) -- ++(0, -2.25) -| (ae);
		
		\begin{pgfonlayer}{background}
			\node (shade) [draw=red, fill=red!10, fit=(res1)(res4)(res6)(dot1), inner sep=5pt, dashed, rounded corners] {};
		\end{pgfonlayer}
		
	\end{tikzpicture}
	\caption{Architectural overview of the latent block, where modified components are highlighted in red.}
	\label{fig:latent_block}
\end{figure}
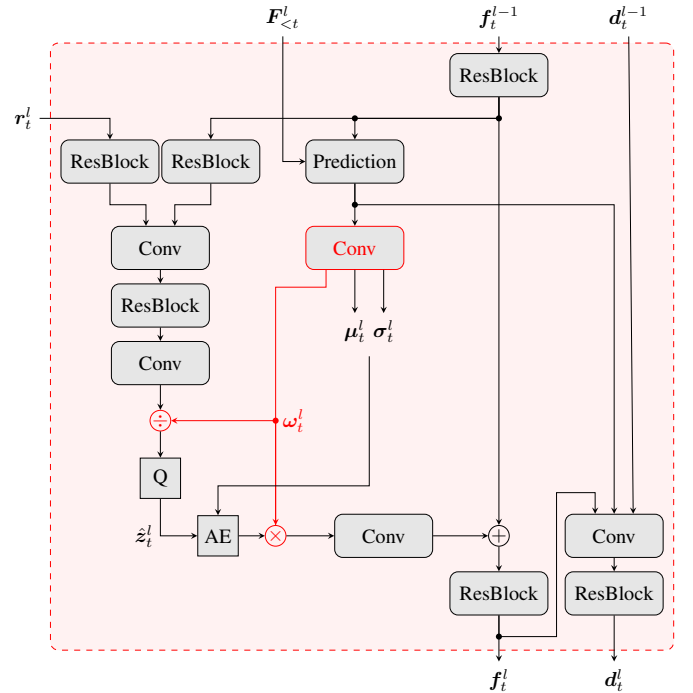

MR-MD-SSF~\cite{fathima2023} combines a global variable-rate approach with a pixel-accurate binary ROI map, where the relevance of the distortion loss for non-ROI pixels is decreased.
This work builds upon the Scale Space Flow model~\cite{agustsson2020} where the ROI mask is concatenated to the inputs of the I-frame encoder, the flow encoder and the residual coder.
Using this input, the model learns to perform adaptive spatial rate allocation.

Both designs are limited in terms of spatial rate allocation.
For ROI-DVC, the rate-allocation is constrained to the fixed spatial size of the segmented patches, and the rate point options for a specific patch are limited by the number of trained models.
While MR-MD-SSF allows higher accuracy in the rate allocation process via the pixel-accurate mask, the design is constrained by the binary input, which splits pixels into either ROI or non-ROI.
Thus, there is still a need for pixel-based fine-granular local rate control in neural video compression.

\section{Novel Approach for Local Rate Control in Neural Video Compression}
We base our approach on the deep hierarchical video compression network DHVC~\cite{lu2024}.
In contrast to many other neural video compression networks, which require a separately transmitted I-frame as initial reference for all subsequent frames, DHVC is applicable for both I- and P-frames.
This is achieved via temporal feature buffers, which are initially filled with learned temporal biases and updated throughout the coding process.
DHVC uses a hierarchical latent design, where the information is transmitted on various scales.
Furthermore, DHVC applies implicit motion modeling and, therefore, does not require motion estimation and compensation between frames.

\subsection{Local Rate Control}
Our goal is to develop a continuous local rate control option for neural video compression via a pixel accurate mask.
For this purpose, we concatenate a quality map $\bm{m}_t$ with same spatial dimensions to the input frame, as shown in Fig.~\ref{fig:lrc_dhvc}.
This map contains the weights for the rate distortion trade-off for each individual pixel in the video frame.
The value range of the quality map is normalized to $\bm{m}_t \in [0, 1]$, where a value of zero corresponds to lowest rate and quality, and a value of one to the highest ones.

Furthermore, we extend the network branch generating the prior parameters $\bm{\mu}$ and $\bm{\sigma}$ for the Gaussian entropy model inside the latent blocks to derive an additional weight parameter $\bm{\omega}$.
The latent values are adaptively divided by the elements in the weight parameter to scale the latent tensor before quantization.
After arithmetic coding, the latent tensor is rescaled again by multiplying it with the weight parameters $\bm{\omega}$.
The architecture of the latent block is shown in Fig.~\ref{fig:latent_block}.

\subsection{Training Setup}
\label{sec:experiment_design}
\begin{table}[t]
	\centering
	\normalsize
	\begin{tabular}{ccc}
		\hline
		Epoch	&	Dataset		&	Number of frames	\\
		\hline
		25 	 	&	Vimeo90k	&	3	\\
		35		&	Vimeo90k	&	7	\\
		40		&	Vimeo-32	&	16	\\
		50		&	Vimeo-32	&	32	\\
		\hline
	\end{tabular}
	\caption{Training schedule with increasing number of frames.}
	\label{tab:training}
\end{table}
For all experiments, we train the compression model using the Vimeo90k dataset~\cite{xue2019} and a dataset which reprocesses the original Vimeo videos to obtain sequences of 32 frames, to allow training on more than seven frames.
We start the training with short sequences consisting of three frames and increase up to 32 frames per sequence.
The different training steps are shown in Table~\ref{tab:training}.

In order to learn a quality map dependent rate-distortion trade-off, the distortion loss has to incorporate the quality map $\bm{m}_t$.
First, the normalized quality map is rescaled into the lambda domain:

\begin{equation}
	\bm{\Lambda}_t = \alpha \cdot e^{\beta \cdot \bm{m}_t}.
\end{equation}

We empirically derive the parameters \mbox{$\alpha=0.001$} and \mbox{$\beta=6$}, which lead to a reasonable range of lambda values matching common ranges in neural video coding.
The weighted MSE loss is derived by an elementwise multiplication of the elements in $\bm{\Lambda}_t$ and the quadratic error of the pixel values:

\begin{figure}[!t]
	\centering
	\begin{subfigure}{\columnwidth}
		\centering
		\includegraphics[width=\columnwidth]{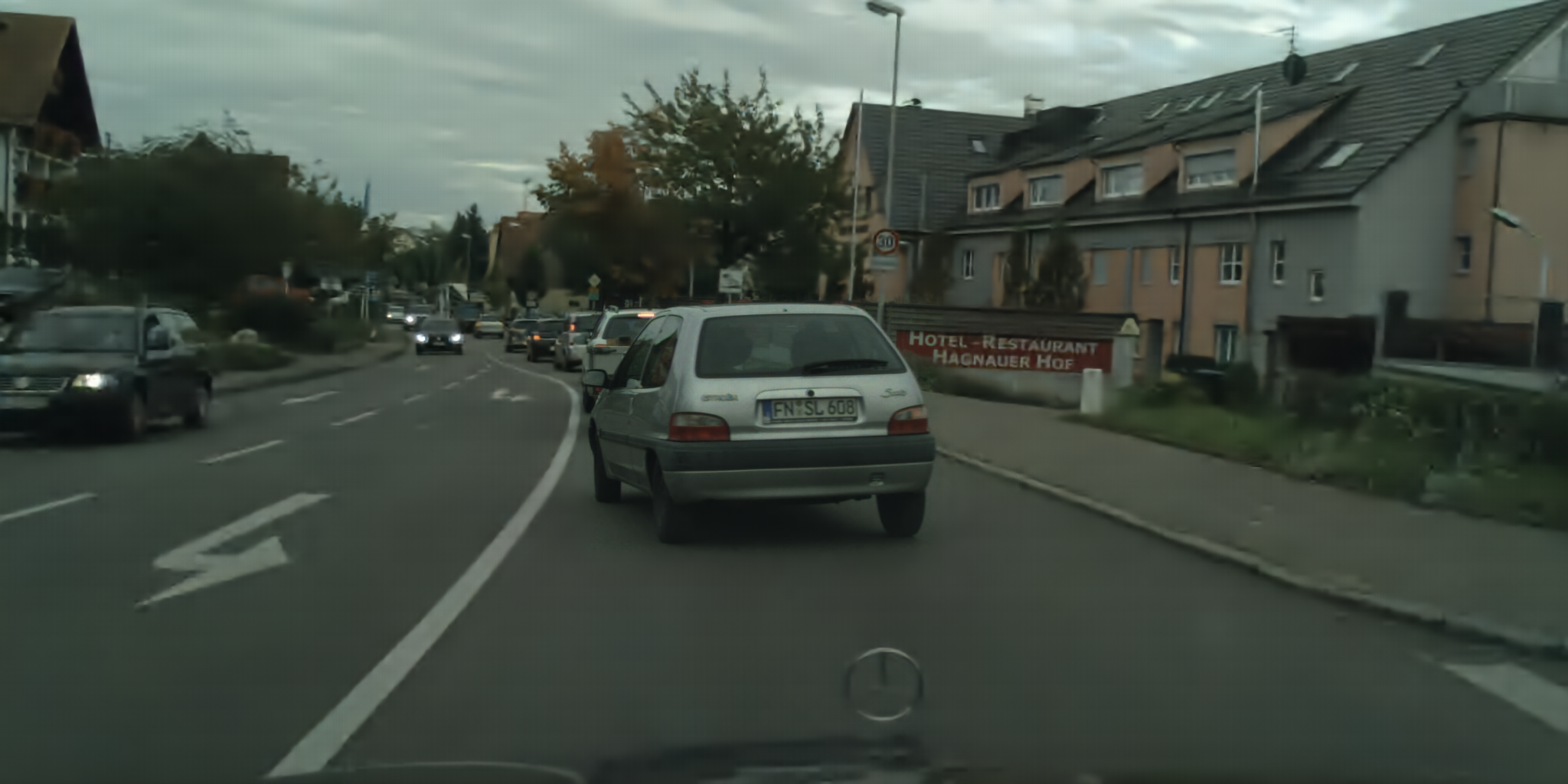}
		\caption{Compressed frame}
		\label{subfig:img}
	\end{subfigure}\\
	\vspace{7pt}
	\begin{subfigure}{\columnwidth}
		\centering
		\includegraphics[width=\columnwidth]{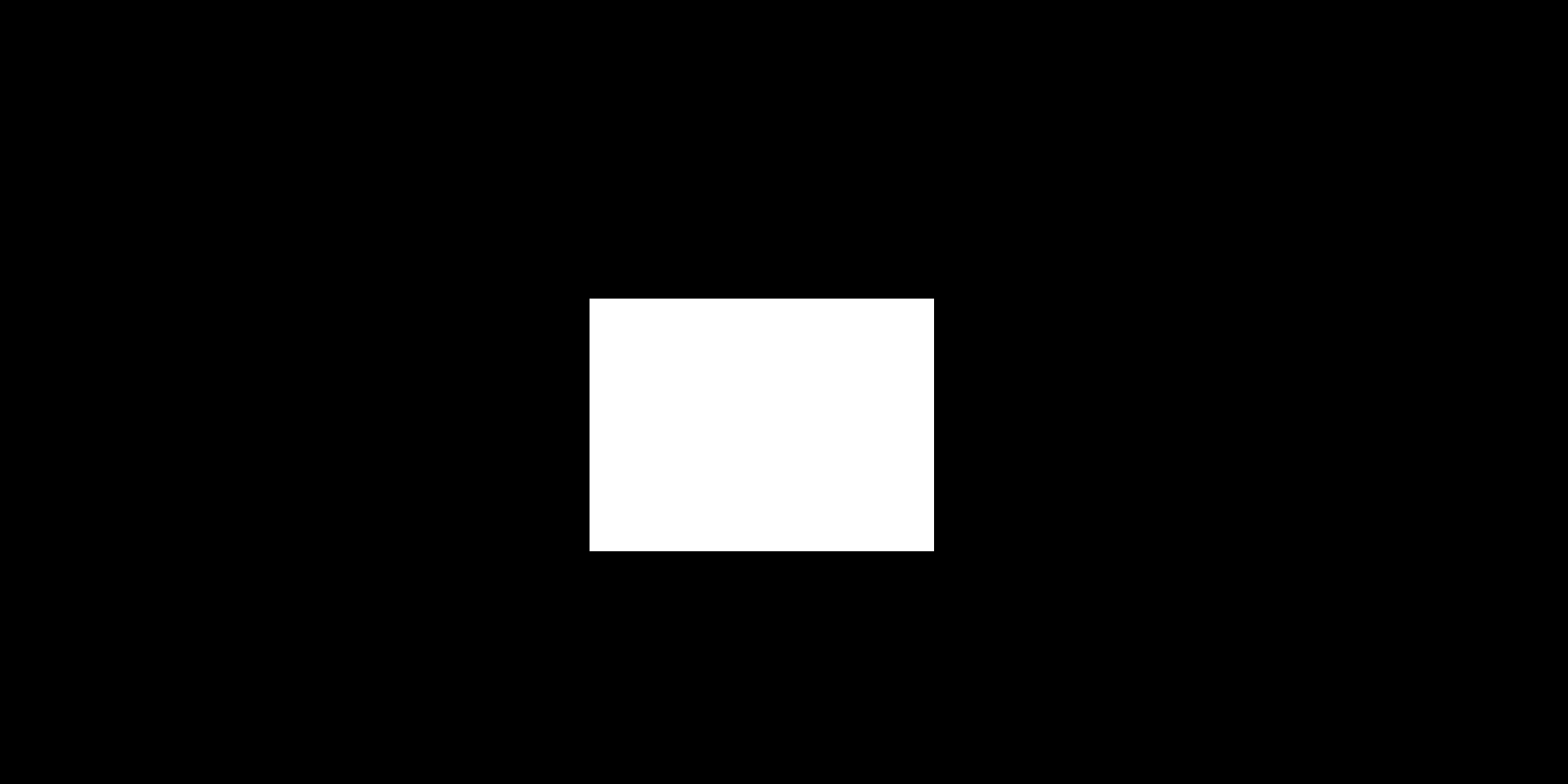}
		\caption{Quality map}
		\label{subfig:qualitymap}
	\end{subfigure}\\
	\vspace{7pt}
	\begin{subfigure}{\columnwidth}
		\centering
		\includegraphics[width=\columnwidth]{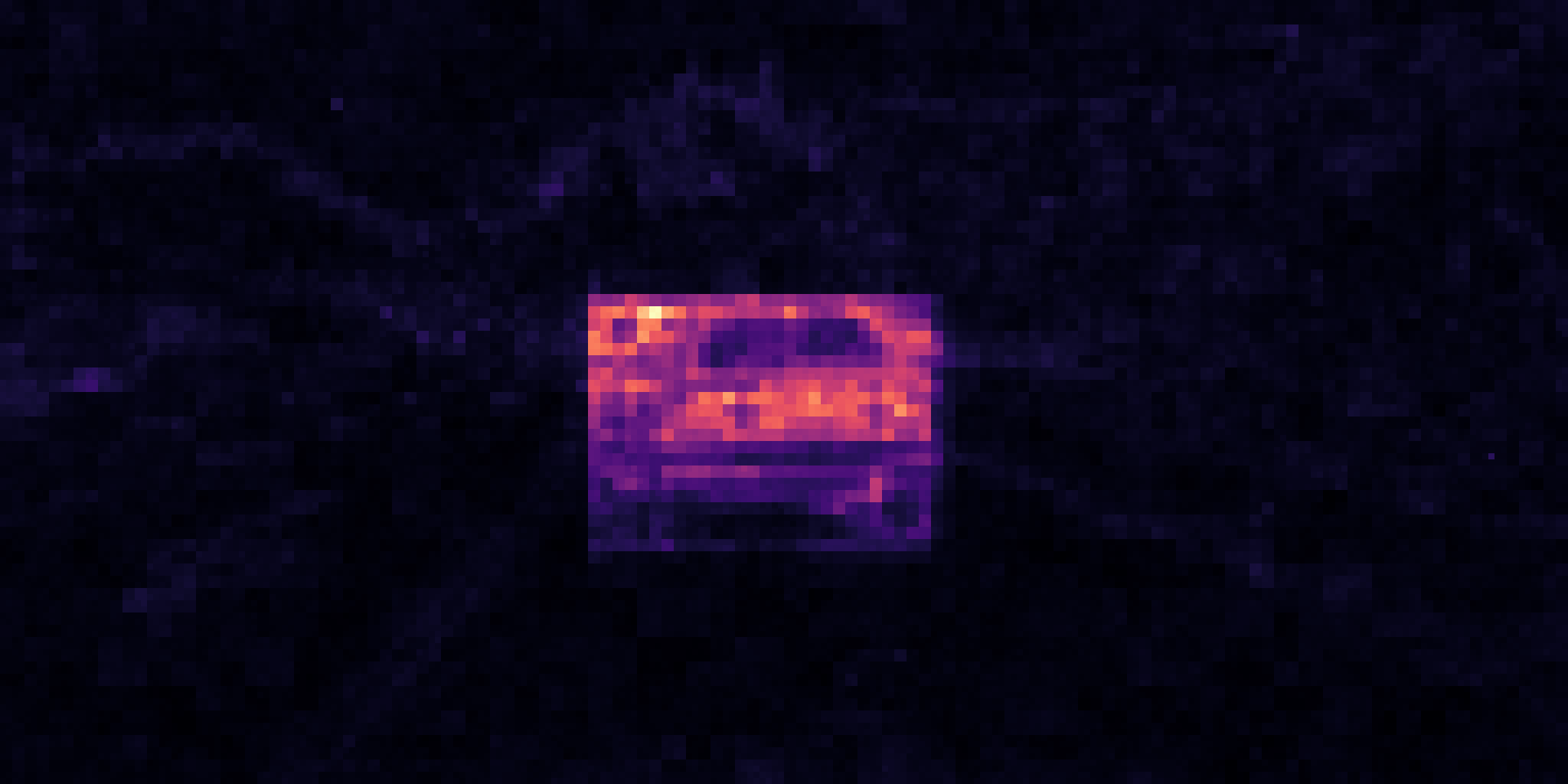}
		\caption{Bit distribution heatmap}
		\label{subfig:heatmap}
	\end{subfigure}\\
	\caption{Exemplary compressed frame with corresponding quality map (white: high quality) and bit distribution heatmap. \textit{Best to be viewed enlarged on a screen.}}
	\label{fig:example}
\end{figure}
\begin{equation}
	\L_{\text{W-MSE}} = \frac{1}{H \cdot W} \cdot \sum_{H, W} \bm{\Lambda}_t \odot (\bm{x}_t - \bm{\hat{x}}_t)^2.
	\label{eq:roi_loss}
\end{equation}

Pixels with a higher value in the quality map $\bm{m}_t$ correspond to a larger lambda value and, thus, have a higher influence on the total loss.
By using an exponential basis function for the mapping from the quality map values to the lambda values $(\bm{m}_t\rightarrow\bm{\Lambda}_t)$, we ensure a wide bit rate range and a reasonable spacing when sampling uniformly from the quality range $[0, 1]$.
The total loss is derived by the addition of the estimated bit rate and the weighted MSE loss:

\begin{equation}
	\L = \L_{\text{rate}} + \L_{\text{W-MSE}}
\end{equation}

To obtain diverse quality maps during training, we generate them in a constrained-random fashion.
The quality maps can contain diverse features, like quality fading, sharp quality boundaries, and nearly homogeneous regions.
Furthermore, we ensure temporal stability throughout multiple frames by updating the previous quality map instead of generating independent quality maps for each frame.
This is done by placing random pseudo-objects with random quality values in the initial quality map.
For consecutive quality maps, the position of an pseudo-object can only move by a limited amount of pixels.
The change in the quality value for a pseudo-object is limited as well.

\section{Qualitative Analysis of Local Rate Allocation and Rate-Distortion Results}
\begin{figure}[tb]
	\centering
	% This file was created with tikzplotlib v0.10.1.
\begin{tikzpicture}
	
	\definecolor{darkgray176}{RGB}{176,176,176}
	\definecolor{goldenrod18818934}{RGB}{188,189,34}
	\definecolor{orange}{RGB}{255,165,0}
	\definecolor{steelblue}{RGB}{70,130,180}
	
	\begin{axis}[
		tick align=outside,
		tick pos=left,
		x grid style={darkgray176},
		xlabel={Rate in bpp},
		xmajorgrids,
		xmin=0, xmax=0.7,
		xtick style={color=black},
		y grid style={darkgray176},
		ylabel={PSNR in dB},
		ymajorgrids,
		ymin=30, ymax=45,
		ytick style={color=black},
		legend pos={south east}
		]
		\addplot [semithick, black, mark=*, mark size=2.5, mark options={solid}, dashed]
		table {%
			0.04122496337890632 36.69418545662694
			0.06284843343098963 37.960945746483894
			0.09960460408528643 39.2407402813078
			0.15023156026204407 40.221538436655756
		};
		\addlegendentry{DHVC \cite{lu2024}}
		\addplot [semithick, steelblue]
		table {%
			0.03375435078938802 33.221475032482104
			0.039184176635742196 33.77408264226545
			0.0458095286051432 34.34504049827289
			0.053705330403645835 34.91550547067989
			0.063103423055013 35.48462038730205
			0.07455343119303381 36.06834940565267
			0.0883552134195965 36.6583644785375
			0.10503991495768229 37.258549737037896
			0.12499732971191395 37.85008287030901
			0.14597179565429694 38.36707508142211
			0.1698524098714193 38.8633656032657
			0.19707544352213532 39.33648658090205
			0.2280716481526692 39.7855255825031
			0.2631628896077474 40.20800154166265
			0.30314928181966144 40.60589259555108
			0.34874365437825544 40.97739764632833
			0.4006111236572266 41.32031072241663
			0.45937609354654957 41.63280258878704
			0.5256579467773441 41.91344082173892
			0.5999240437825526 42.16091250858258
			0.6818828989664705 42.3736706916316
		};
		\addlegendentry{LRC-DHVC (Uniform)}
		\addplot [semithick, orange]
		table {%
			0.03396478983561193 34.64946055716711
			0.04884088033040362 36.00637700409164
			0.06964976298014323 37.109310297799446
			0.1022518910725911 38.2640205997631
			0.1414561655680338 39.34626321405903
			0.20376520792643246 40.41170874433622
			0.2898218475341795 41.42666412795811
			0.39829593709309924 42.01075143584106
		};
		\addlegendentry{LRC-DHVC (Optimized)}
	\end{axis}
	
\end{tikzpicture}
	\caption{Rate-distortion curves comparing single-rate models for DHVC~\cite{lu2024} with our LRC-DHVC which integrates local rate control in a single model. The continuous RD-curve of LRC-DHVC (Uniform) is approximated by the evaluation of 21 rate points ($\bm{m}_t=\{0, 0.05, ..., 0.95, 1.0\}$). The eight distinct quality maps for LRC-DHVC (Optimized) are derived from the hyperparameter set \mbox{($\lambda=\{64, 128, 256, 512, 1024, 2048, 4096, 8192\}$)}.}
	\label{fig:dhvc_results}
\end{figure}
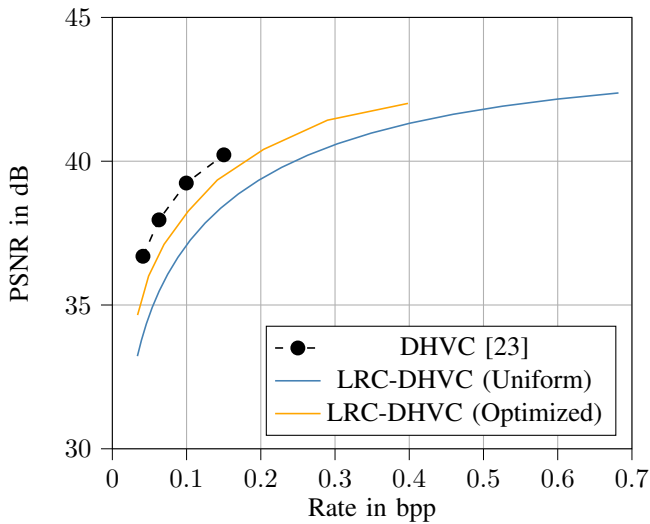
We use the video sequences in the Cityscapes~\cite{cordts2016} dataset to test our compression network on previously unseen data.
In order to test the lower and upper bound of bit rate and quality of our local rate control coder, we evaluate the network using homogeneous quality maps.
With the limits of $\bm{m}_t=0$ and $\bm{m}_t=1$, we achieve a diverse bit rate and quality range spanning from (0.0338~bpp, 33.22~dB) up to (0.6819~bpp, 42.37~dB).

Fig.~\ref{fig:example} shows an examplary compressed frame (a), the applied quality map (b) and the bit distribution heatmap (c).
The quality map was chosen as a rectangluar shape around the car in the center.
For the high quality region, we obtain 43.66~dB compared to 30.34~dB for the surrounding background area.
Both, the quality in the compressed image and the bit distribution heatmap show that LRC-DHVC captures the information in the quality map and focuses on the salient region.

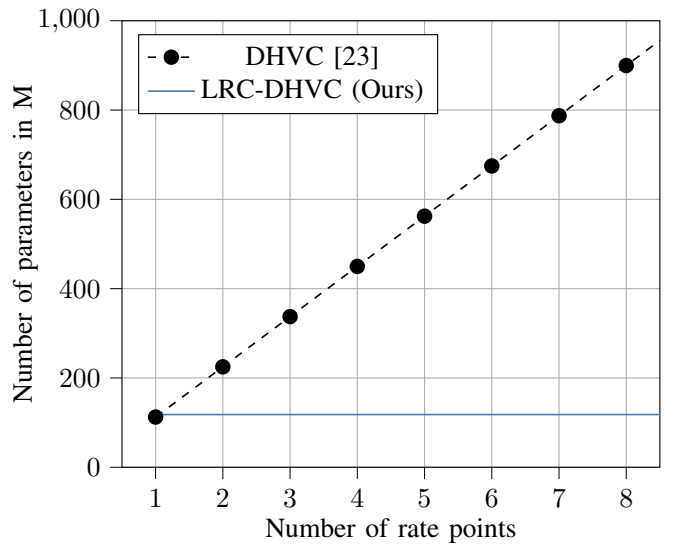
\begin{figure}[tb]
	\centering
	% This file was created with tikzplotlib v0.10.1.
\begin{tikzpicture}
	
	\definecolor{darkgray176}{RGB}{176,176,176}
	\definecolor{goldenrod18818934}{RGB}{188,189,34}
	\definecolor{orange}{RGB}{255,165,0}
	\definecolor{steelblue}{RGB}{70,130,180}
	
	\begin{axis}[
		tick align=outside,
		tick pos=left,
		x grid style={darkgray176},
		xlabel={Number of rate points},
		xmajorgrids,
		xmin=0.5, xmax=8.5,
		xtick style={color=black},
		xtick distance={1},
		y grid style={darkgray176},
		ylabel={Number of parameters in M},
		ymajorgrids,
		ymin=0, ymax=1000,
		ytick style={color=black},
		legend pos={north west}
		]
		\addplot [semithick, black, mark=*, mark size=2.5, mark options={solid}, dashed]
		table {%
			1 112.46
			2 224.92
			3 337.38
			4 449.84
			5 562.30
			6 674.76
			7 787.22
			8 899.68
			9 1012.14
		};
		\addlegendentry{DHVC \cite{lu2024}}
		\addplot [semithick, steelblue]
		table {%
			1 118.02
			2 118.02
			3 118.02
			4 118.02
			5 118.02
			6 118.02
			7 118.02
			8 118.02
			9 118.02
		};
		\addlegendentry{LRC-DHVC (Ours)}
	\end{axis}
	
\end{tikzpicture}
	\caption{Relation between number of rate points and number of parameters for DHVC~\cite{lu2024} and LRC-DHVC. The parameter count for LRC-DHVC remains constant, whereas it increases linearly for DHVC.}
	\label{fig:rate_points}
\end{figure}

LRC-DHVC can compress videos within wide quality and bit rate ranges.
In Fig.~\ref{fig:dhvc_results}, we compare the rate-distortion curves of our local rate control enabled LRC-DHVC using uniform quality maps with the baseline DHVC, which uses four distinct single-rate point models.
Using uniform quality maps with LRC-DHVC leads to a BD-rate increase by 99~\% compared to DHVC~\cite{lu2024}.
Uniform quality maps, however, are by far not optimal, even if uniform quality is wanted, since rate allocation depends not only on the desired quality but also on the scene complexity of an area within a video frame.

Therefore, we derive optimized quality maps with respect to the rate-distortion trade-off during inference.
These optimized quality maps significantly outperform the naive approach of using uniform quality maps.
The gap to the single-rate model diminishes to a BD-rate increase of 41~\%.

\begin{table}[t]
	\centering
	\normalsize
	\begin{tabular}{ccc}
		\hline
		& kMAC/pix	&	Mparam	\\
		\hline
		DHVC~\cite{lu2024} 	& 446.23	&	117.9	\\
		LRC-DHVC (ours)		& 446.85	&	118.1	\\
		\hline
		rel. inc.			& 0.13 \%	&	0.15 \%	\\
		\hline
	\end{tabular}
	\caption{Complexity evaluation for compression of a frame of size $1024\times 2048$ in multiply-accumulate operations per pixel and parameter count for one model.}
	\label{tab:complexity}
\end{table}

The major benefit and designed use-case of LRC-DHVC, however, lies within the locally-varying and continuous rate control for neural video compression over a large quality range within a single model.
Fig.~\ref{fig:rate_points} shows the relation between the number of rate points and the parameter count for LRC-DHVC and DHVC.
While the number of parameters for LRC-DHVC remains constant, it increases linearly with each additional rate point for DHVC.
In order to span the same quality range of LRC-DHVC, the baseline DHVC would require approximately eight rate points, which would lead to a total parameter count of $\sim$0.9~Billion, as opposed to $\sim$118~Million for our approach.
Table~\ref{tab:complexity} compares the compression complexity of the single-rate DHVC~\cite{lu2024} with LRC-DHVC.
LRC-DHVC offers diverse local and global rate points at a relative parameter increase of 0.15~\% compared to one single-rate DHVC model.
The modifications to offer local rate control require an insignificant increase of compression complexity of 0.13~\%.

\section{Conclusion}
In this paper, we propose LRC-DHVC, a unified compression model which achieves a wide range and locally-varying rate and quality distribution for neural video compression, by concatenating a continuous quality map and applying a spatially-weighted distortion loss.

Our experiments demonstrate that LRC-DHVC is able to accurately follow the desired rate distribution enabling spatial rate allocation across a wide quality range within a single model.
Thus, it is promising to apply the compression network to use cases which benefit from region-specific rate-distortion trade-offs, such as video coding for machines.

Task-adaptive quality maps for LRC-DHVC can be generated with our related work~\cite{windsheimer2026}
While LRC-DHVC with continuous local rate control cannot fully reach the compression performance of similar sized single rate compression networks, we still see the potential to close this gap in further research.
Additionally, this concept can be applied to a variety of other video compression networks, such as the latest versions of the DCVC family~\cite{li2024}, \cite{jia2025}.

%\vfill\pagebreak
%\ninept

% -------------------------------------------------------------------------
\bibliographystyle{IEEEbib}
\bibliography{refs}

\end{document}